# On-chip time resolved detection of quantum dot emission using integrated superconducting single photon detectors

G. Reithmaier[†], S. Lichtmannecker[†], T. Reichert[†], P. Hasch[†], K. Müller[†], M. Bichler[†],

R. Gross[‡] and J. J. Finley[†]

[†]*Walter Schottky Institut, Physik Department and Center of Nanotechnology and Nanomaterials, Technische Universität München, Garching, 85748, Germany*

[‡] *Walter Meißner Institut, Technische Universität München, Garching, 85748, Germany*

We report the routing of quantum light emitted by self-assembled InGaAs quantum dots (QDs) into the optical modes of a GaAs ridge waveguide and its efficient detection on-chip via evanescent coupling to NbN superconducting nanowire single photon detectors (SNSPDs). Individual QD light sources embedded within such integrated nano-photonic circuits are highly attractive for the realization of quantum photonic circuits for many applications in photonic information science. Here, we demonstrate that the waveguide coupled SNSPDs primarily detect QD luminescence with scattered photons from the excitation laser being negligible by comparison. The SNSPD detection efficiency from the evanescently coupled waveguide modes is shown to be two orders of magnitude higher when compared with operation under normal incidence illumination. Furthermore, in-situ time resolved measurements show an average exciton lifetime of 0.93 ± 0.03 ns when recorded with the integrated detector with an ultrafast timing jitter of only 72 ± 2 ps showing the great potential of this highly integrated quantum optics system.

_________________________________________________________________

To whom correspondence should be addressed:

E-mail: guenther.reithmaier@wsi.tum.de and finley@wsi.tum.de

KEYWORDS: superconducting single photon detectors, SNSPDs, quantum dots, integrated quantum nanophotonics

Photonic information technologies using semiconductors are ubiquitous and are rapidly being pushed to the *quantum limit* where non-classical states of light can be generated and manipulated in nanoscale optical circuits.[1,2] Single photons can be readily generated on-chip[3] and preferentially routed into waveguide modes by carefully tailoring the local density of photonic modes experienced by the emitter[4,5,6]. Furthermore, *effective* interactions between photons can be induced by exploiting coherent light-matter couplings between the tightly localized vacuum field in nanoscale cavities, leading to remarkable phenomena such as *photon blockade*[7,8] needed for optical transistors[9,10] and ultrafast optical switching with only a few photons[11]. While the generation and routing of quantum light on a semiconductor chip[3,12,13] has already been demonstrated by several groups, the ability to generate *and* detect single photons on-chip with near unity quantum efficiency[14] and, moreover, integrate sources and detectors with nanophotonic hardware such as waveguides, high-Q nanocavities and beamsplitters would represent a major step towards the realization of semiconductor based quantum optical circuits.

In SNSPDs, photon detection occurs via the formation of a normal conducting hotspot in a thin superconducting nanowire upon the absorption of a single photon.[15] Since the bias current flowing through the nanowire is slightly sub-critical ($\sim 0.95\ I_{crit}$), the local heating arising from single photon absorption results in the breakup of Cooper pairs, local switching of the nanowire to a normal conducting state and a measurable voltage pulse in the external readout circuit. Such SNSPDs provide very high single photon detection efficiencies[16,17,18,19,14], low dark count rates[20], sensitivity from the visible to the IR[21] and picosecond timing resolution[22,23]. The possibility to integrate superconducting nanowire single photon detectors (SNSPDs) onto dielectric[4,14] and plasmonic[5] waveguides results in a drastic increase of the absorption length for incoming photons, pushing the single-photon detection efficiency towards unity. Both the generation of cluster states of photonic qubits for one-way quantum computation[24] and the measurement based teleportation schemes[25] rely on having such near perfect detection efficiency.

Here, we demonstrate the on-chip generation of light originating from optically pumped micro-ensembles of ~ 120 self-assembled InGaAs QDs, low loss guiding over ~ 0.5 mm along a GaAs-AlGaAs ridge waveguide and high efficiency detection via evanescent coupling to an integrated SNSPD. By comparing measurements performed with optical excitation above and below the GaAs bandgap and exploring the temporal response of the system, we show that the detector signal overwhelmingly stems from QD luminescence with a negligible background from the laser. Power dependent measurements confirm the single photon sensitivity of the detectors and show that the SNSPD is about two orders of magnitude *more sensitive* to waveguide photons than when illuminated in normal incidence. In-situ time resolved measurements performed using the integrated detector show an average QD spontaneous emission lifetime of 0.93 ± 0.03 ns, with a low timing jitter of only 72 ± 2 ps. The performance metrics of the SNSPD integrated directly onto GaAs nano – photonic hardware confirms the strong potential for on-chip few-photon quantum optical experiments on a semiconductor platform[2].

The samples investigated were grown using solid source molecular beam epitaxy and consisted of a 350 μm thick GaAs buffer onto which a 2 μm thick $Al_{0.8}Ga_{0.2}As$ waveguide cladding layer was deposited. Following this, a 250 nm thick GaAs waveguide core was grown into which a layer of self-assembled InGaAs quantum dots was embedded at its midpoint. The growth conditions used resulted in dots with a typical lateral (vertical) size of 25 ± 5 nm (5 ± 1 nm) as shown in the inset of fig 1a, an areal density of 6 ± 1 $μm^{-2}$ and photoluminescence emission around ~ 920 nm at 4 K with a FWHM of 60 nm. After growth, the native oxide was removed from the sample surface using an HCl dip and a high quality 10 ± 0.5 nm thick NbN superconducting film was deposited using DC reactive magnetron sputtering. By carefully optimizing the deposition temperature, rate and the Nb:N ratio, high quality superconducting films were obtained on the GaAs substrate ($T_C$ = 10.2 ± 0.2 K), despite the 26% lattice mismatch[26,27,19]. The nanowire detector was then defined using electron beam lithography with a negative tone resist and reactive ion etching using a $SF_6$ / $C_4F_8$ plasma to form an NbN nanowire meander consisting of 34x, 80 ± 10 nm wide nanowires separated by 170 ± 10 nm to form a detector with a total length of 23 μm along the waveguide axis and a width of 8.5 μm. A scanning electron microscope image of the resulting NbN nanowires on GaAs is presented in figure 1a - inset.

Subsequently ~ 500 μm long, 17 μm wide multimodal ridge waveguides were defined using photolithography and wet etching in a citric acid + $H_2O_2$ solution. The waveguides feature a 90° gradual bend at their midpoint having a radius of curvature of 150 μm as depicted schematically in fig 1a and the SNSPD threads one end of the waveguide. Figure 1b shows the layer sequence of the epitaxial layers and a vertical refractive index profile through the waveguide cladding, core and superconducting NbN nanowire. In order to estimate the maximum detection efficiency of such SNSPDs we simulated the optical field distribution of the fundamental waveguide mode using a commercial-grade eigenmode solver and propagator.[28] The results obtained clearly show the maximum optical intensity close to the QD layer, as shown by the contour plot in figure 2b. Using the measured dielectric function of the NbN film ($\epsilon = 3.5 + i\,3.8$) we calculated that 97.8% of incident waveguide photons are absorbed by the detector, in good agreement with recent findings[14,29] for passive waveguide integrated SNSPDs that revealed detection efficiencies close to unity.

Spatially resolved photoluminescence (PL) measurements were performed whilst the sample was held at a nominal temperature of 4.2 K inside a cryogenic microwave probe station with optical access. This system provides a diffraction limited laser spot with a diameter of ~ 5 µm and allows the SNSPD to be contacted using GHz voltage probes, thus, facilitating in-situ detection of PL routed along the waveguide. Data recorded using this measurement system is termed *on-chip PL* in the discussion below. The SNSPD was operated using a bias-tee to drive a fixed bias current of $I = 0.95 \times I_C = 6.0 \, \mu A$ through the nanowires[19]. Voltage pulses arising from single photon detection events were then amplified and detected with a 350 MHz frequency counter. For the chosen operation conditions the 10 nm thick NbN SNSPD shows a negligible dark count rate < 10 cps and a top-illumination detection efficiency of 0.001% for light at 940 nm, as expected for the relatively thick 10 nm NbN film[21]. As discussed below, the detection efficiency for *waveguide photons* is about two orders of magnitude larger due to significantly longer interaction length. Additional PL-spectroscopy measurements were performed on the same sample with excitation and detection normal to the waveguide axis using low temperature confocal microscope with a much higher spatial resolution (~ 1 µm). PL-spectra are obtained by dispersing the emitted light using a 0.5 m imaging monochromator and detected using a silicon CCD detector. In the following, such measurements are termed *confocal-PL*.

We begin by discussing *on-chip PL* recorded by raster scanning the excitation laser spot across the entire active waveguide structure. Typical results are presented in fig 2a that compares false color images of the SNSPD count rate recorded using an excitation wavelength above the GaAs bandgap ($\lambda = 632.8 \, nm$ - red color coding) and selected regions of the device mapped with much longer wavelength excitation, far below the GaAs bandgap ($\lambda = 940$ nm - blue color coding). All waveguide scans in fig 2a were recorded using the same excitation power density of 25 W/cm². At 940 nm, the excitation efficiency of the QDs is expected to be $\sim 10^4 \times$ lower[30] whilst the reduction of the SNSPD sensitivity is much weaker ($\sim 3 - 6 \times$)[21]. Using above gap excitation the form of the waveguide can clearly be identified in the on-chip PL map in fig 2a, the count rate increasing significantly as the laser spot moves closer to the detector. In strong contrast, using 940 nm excitation only a background of $\sim 1000$ cps is observed with no visible signal enhancement as the laser spot is scanned onto the waveguide. To systematically probe the wavelength selectivity we performed line scans across the remote waveguide end, farthest from the SNSPD, along the line A-B marked in fig 2b. Line scans were made using non-resonant excitation above the GaAs bandgap (632.8 nm), below the GaAs bandgap into the wetting layer continuum (830 nm) and resonantly into the s-shell transitions of the QDs using far below bandgap excitation (940 nm). For above gap and wetting layer excitation, the waveguide can clearly be distinguished in the line scans. In contrast, with 940 nm excitation the waveguide topology could *not* be imaged (see fig 2b and fig 2a-inset) despite the detector remaining highly sensitive to such IR-illumination. This expectation is confirmed by the data presented in the inset of fig 2a that shows the direct normal incidence response of the detector at 940 nm when raster scanning the laser spot with a low power density of 0.4 W/cm². A clear maximum is observed when the laser spot is incident on the detector and, by measuring the normal incidence count rate and carefully calibrating the incident photon flux onto the detector, we estimated the top-illumination quantum efficiency to be $\sim 0.001\%$ at 940 nm, in good accord with previous measurments[19,21] and expectations for a 10 nm thick NbN film. These observations clearly indicate that the signal detected when exciting on the waveguide arises from QD PL emitted into the waveguide mode and guided to the SNSPD whereupon it is evanescently absorbed by the SNSPD. This conclusion is unequivocally validated by the time-resolved measurements presented below.

For excitation at 632.8 nm close to the waveguide bend (fig 2a) using a power density of 25 W/cm², the typical maximum count rate on the detector ranged from 300 – 100 kcps when scanning the laser along the C – D marked line with a background count rate of < 10 kcps recorded at position-F, originating from scattered light on the sample surface. By exponentially fitting the intensity as a function of distance from the detector for the trajectories marked C – D and D – E, the waveguide losses are determined to be 0.022 dB/µm within the bend and 0.005 dB/µm in the straight segments, respectively (analysis in supplementary). Due to the low quantum dot density of ~ 6 µm$^{-2}$, reapsorption by quantum dots along the waveguide can be neglected[31]. As shown later in the discussion of fig 4b, a lower limit for the SNSPD detection efficiency for evanescently coupled quantum dot emission is estimated using these losses.

To unambiguously prove that the detected signal in in-situ PL measurements does indeed stem from QD emission, with a negligible laser background, we used the SSPD to perform time-resolved measurements. Here, the sample was excited using a 653 nm pulsed laser diode focused close to the remote waveguide end ~ 0.5 mm from the SNSPD. This source provided sub 60 ps duration pulses at a repetition rate of 20 MHz with low timing jitter < 3 ps. The SNSPD response was then read-out using a 20 GHz-sampling oscilloscope to record a histogram of the time intervals between the trigger signal provided by the laser diode and the photon detection voltage pulse registered by the detector. Figure 3 shows typical time resolved data on a logarithmic scale including exponential fits to the rising and falling edge of the spontaneous emission dynamics. For comparison the instrument response function (IRF) of the detector and associated electronics was recorded using an IR pulsed laser source (952 nm, < 60 ps pulse duration) focused directly onto the detector to *avoid* excitation of QD PL. Fig 3 shows the temporally sharp IRF from which a low timing jitter of 72 ± 2 ps was obtained (fig 3 – inset)[22]. As jitters < 20 ps have been reported[22] for similar SNSPDs, in this case the jitter is most likely limited by the pulse duration of the laser. We fitted the rise ($t_0$) and fall ($t_1$) times of the observed on-chip PL time transient obtaining values of $t_0$ = 136 ± 21 ps and $t_1$ = 0.93 ± 0.03 ns, respectively. While $t_1$ compares very well to the known spontaneous emission lifetimes of InGaAs QDs[20,21], the surprisingly slow rise-time reflects the timescale for carrier thermalisation and capture into the dots from the surrounding GaAs, demonstrating the clear presence of a phonon bottleneck[32]. Evidently, the peak excitation power density provided by the pulsed excitation source (~ 25 W/cm²) is sufficiently low such as to keep the excitation regime firmly in the *single exciton* limit. In this case, significant free carrier populations are not present in the wetting layer and GaAs matrix that typically result in faster carrier capture and intra-dot carrier relaxation dynamics[32]. This expectation is supported by CW power dependent measurements presented below. Importantly, we note that a fast transient with the temporal profile of the excitation laser pulse is not observed close to t = 0ns in our time resolved measurements illustrating that the detected signal is dominated by QD PL and the SNSPD signal does not contain scattered laser light.

Finally, we estimate a lower limit for the efficiency of our SNSPD by comparing the excitation power dependence of the QD-luminescence signal detected using *on-chip PL*, from ~ 100 dots within the 5 μm diameter laser focal volume[33] and *confocal-PL* from individual dots. A typical spectrum obtained using the confocal-PL geometry with excitation at 632.8 nm and a power density of 6.4 W/cm² is presented in fig 4a. Typically ~ 6-20 sharp lines are observed arising from different individual dots within the ~ 1 $\mu m^2$ laser focal area. The inset in fig 4a shows the spatial distribution of the emission intensity within a 0.1 nm wide wavelength window centered on one such prominent single exciton transition, labeled QD-1 at 940.2 nm. The image clearly shows several localized emission centers arising from individual dots.[33] Figure 4b (open squares) shows the power dependent intensity of QD-1 recorded using confocal-PL. A perfectly linear increase of the intensity is observed for excitation power densities < 10 $W/cm^2$ confirming that the peak QD-1 arises from a single exciton transition from an individual dot.[34] As the excitation power increases to $P_0 \sim 55$ W/cm² the intensity of the single exciton saturates, and is then expected to reduce at higher power as the dot occupation shifts further into the multi-exciton regime. The maximum PL-intensity arising from the neutral exciton transition corresponds to a time averaged exciton occupation probability in the dot of $N_X \sim 1$.[34] The majority of the emission lines observed in fig 4a exhibit a similar power dependence indicating that, for the low excitation power densities used in fig 4a, they all stem from single exciton transitions from different dots addressed by the laser. Assuming that the number of randomly generated electron hole pairs inside a dot at any instant obeys Poisson statistics, an assumption that is likely to hold for excitation power densities $P \ll P_0$, [35] the probability that any dot is occupied with a single exciton is $P_X = \alpha \times exp(-(P/P_0))$ where $\alpha = P/P_0$ and the probability of $n$-excitons populating the dot is given by $P_{nX} = \alpha^n \times \exp\left(-\frac{P}{P_0\,n!}\right)$. For $P \ll P_0$ the photon emission rate into the waveguide mode by a *single* pumped dot will, to a good approximation, be dominated by single excitons and, hence, is given by $\phi \sim \frac{P_X(P)}{t_1} = \left(\frac{P}{P_0}\right)\exp(-P/P_0)/t_1$, where $t_1$ = 0.93 ns is the spontaneous emission lifetime determined above. Thus, for an excitation power density of ~ 10 W/cm² ($P/P_0 \sim 0.2$, $exp(-P/P_0) = 0.82$), the maximum photon flux generated in the waveguide mode by dots within the laser focal volume is $\phi_{max} \sim \left(\frac{P}{P_0}\right) \exp\left(-\frac{P}{P_0}\right) N_{QD} \, A_{laser} \, \zeta \, \eta \, / \, t_1$, where $N_{QD} \sim 6 \, \mu m^{-2}$ is the areal density of QDs, $A_{laser}$ is the area of the laser spot, $\zeta = -6.5 dB \equiv 22\%$ accounts for the waveguide losses, as calculated in the discussion related to fig 2, and $\eta = 6\%$ is the radiation fraction of the spontaneous emission into the waveguide mode, calculated by assuming each quantum dot is an ideal point dipole emitter[28]. Using this information we are now in a position to estimate a lower limit for the SNSPD detection probability. For on-chip PL we have $A_{laser} = 20 \mu m^2$ from which we estimate that the maximum possible photon flux in the waveguide mode close to the SNSPD to be $\phi_{max} \sim (0.2 \times 0.82 \times 120 \times 0.22 \times 0.06) / (0.93) \, ns^{-1}$ from which we would estimate a count rate of $\sim 0.3 \pm 0.1 \times 10^9$ cps for the on-chip PL measurement with excitation at 632.8 nm with a power density of $\sim 10$ W/cm$^2$ (filled squares - fig 4b). The on-chip PL dataset clearly shows a linear power dependence with an exponent of 1.06 ± 0.03 reflecting single photon sensitivity of the SNSPD[19]. Taking the measured count rates in on-chip PL into account and taking the experimentally determined waveguide losses and the simulated absorption of the detector, we estimate the detection efficiency of the 10 nm thick SNSPD for evanescently coupled light to be $\sim$ 0.1%. This value could be improved significantly by using thinner NbN layers[Error! Bookmark not defined.,19,21] for which near – unity quantum efficiency can be achieved in waveguide coupled detectors. However, we note that the estimated detectivity of $\sim$ 0.1% is $\sim$ 100x larger than that for normal incidence illumination.

In summary, we presented the creation, routing and detection of single quanta of light on a single chip. Using a waveguide coupled SNSPD detector we showed that on-chip PL provides a detection efficiency of $\sim$ 0.1% for evanescently coupled light which gives a signal enhancement of one order of magnitude when compared to confocal PL. Moreover, it was shown that the integrated detector can be employed to perform in situ time resolved measurements at a temporal resolution of 72 ps, thereby revealing an average exciton lifetime of 0.93 ± 0.03 ns and a rise time of 136 ± 21 ps. The great potential of this highly integrated quantum optics system can be fully used when additionally implementing the spectral resolution of SNSPDs[23] or when performing resonant excitation experiments with single quantum dots.

**Acknowledgements**

We gratefully acknowledge D. Sahin, A. Fiore (TU Eindhoven) and K. Berggren, F. Najafi (MIT) and R. Hadfield (Heriot-Watt) and C. Amann for useful discussions and the BMBF for financial support via QuaHL-Rep Project number 01BQ1036, the EU via the integrated project SOLID and the DFG via SFB-B3 and Nanosystems Initiative Munich.

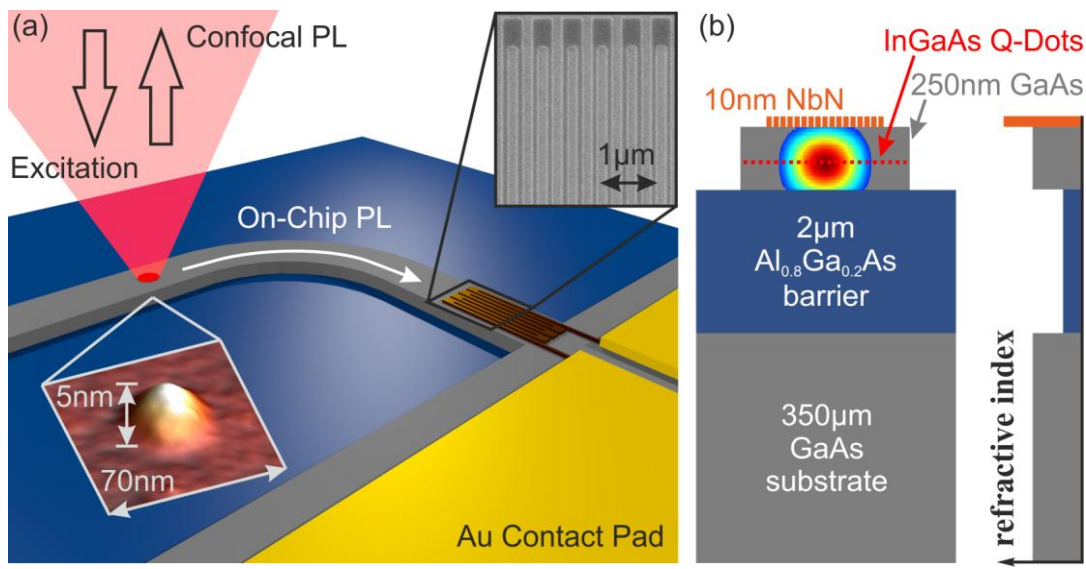

**Figure 1.** (a) Self assembled InGaAs quantum dots, as shown in the AFM image, embedded in a GaAs ridge waveguide are excited using a helium neon laser. The light emitted by the quantum dots is detected either in a confocal geometry or guided along the waveguide and evanescently coupled into a NbN superconducting nanowire single photon detector (SEM image in the inset). (b) Layer structure of the sample as prepared by molecular beam epitaxy and reactive magnetron sputtering. Refractive indices of the materials are schematically depicted on the rightmost side. A single layer of self-assembled InGaAs quantum dots is indicated by the dashed red line, overlapping with the maximum of the simulated intensity of the fundamental waveguide mode, shown in the contour plot.

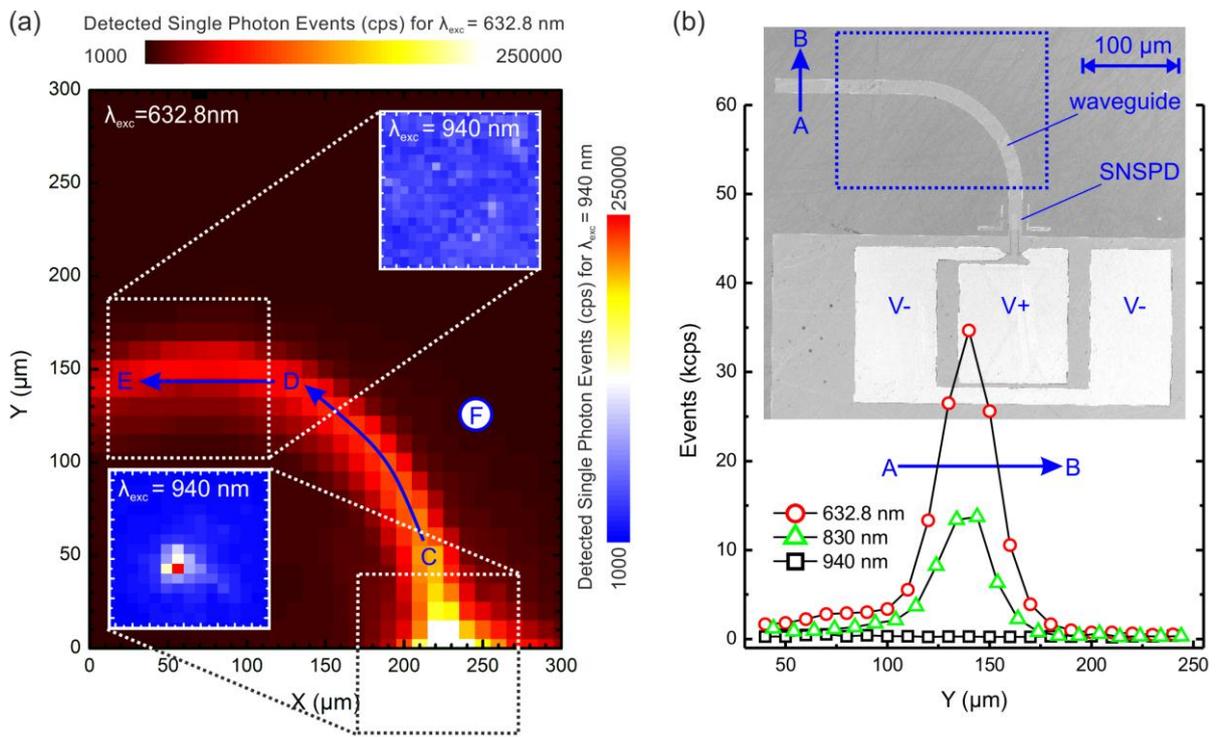

**Figure 2.** (a) Detected single photon events as a function of the laser position along the waveguide for $\lambda_{exc}$ = 632.8 nm, as shown in red colour code. The two insets given with blue colour code each correspond to a (100 µm)² area illuminated with $\lambda_{exc}$ = 940 nm. For illumination at the end of the waveguide only background light is detected, whereas a strong peak is observed at the position of the SNSPD. (b) Recorded single photon events for linescans along the A-B marked line with different excitation

wavelengths. The waveguide can clearly be identified for 632.8 nm and 830 nm, whereas only a small background is detected for 940 nm. The inset shows an SEM image of the analysed strcture with the scan area of fig 2a marked by a dotted line.

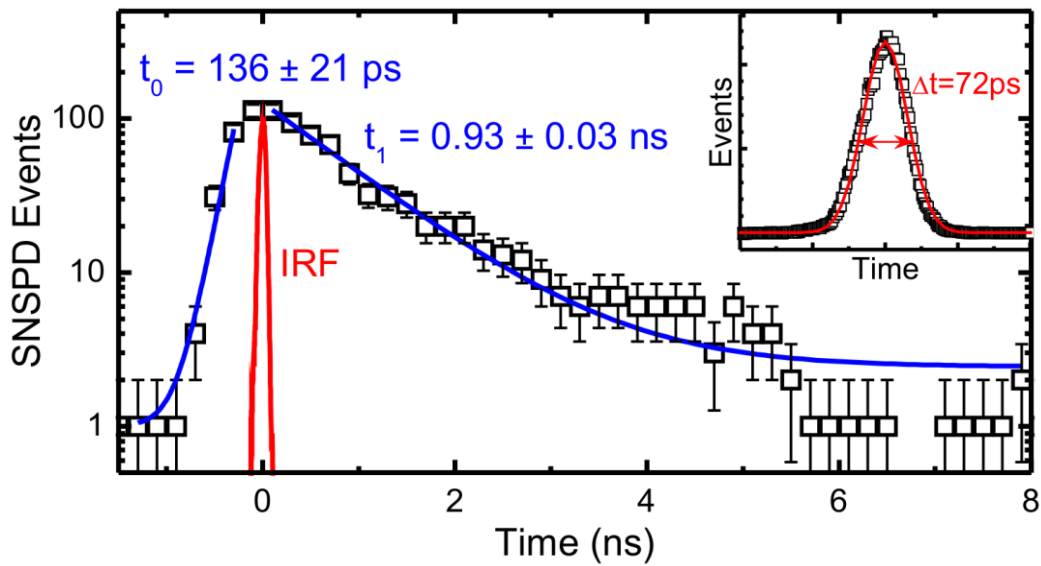

**Figure 3.** SNSPD single photon events for ps - pulsed quantum dot excitation with 653 nm as a function of the photon arrival time shown as black squares. An exponential rise (decay) function given in blue reveals an average photon relaxation time (exciton lifetime) of 136 ± 21 ps (0.93 ± 0.03 ns). The instrument response function showing a temporal resolution of 72 ps is given in red as well as in the inset.

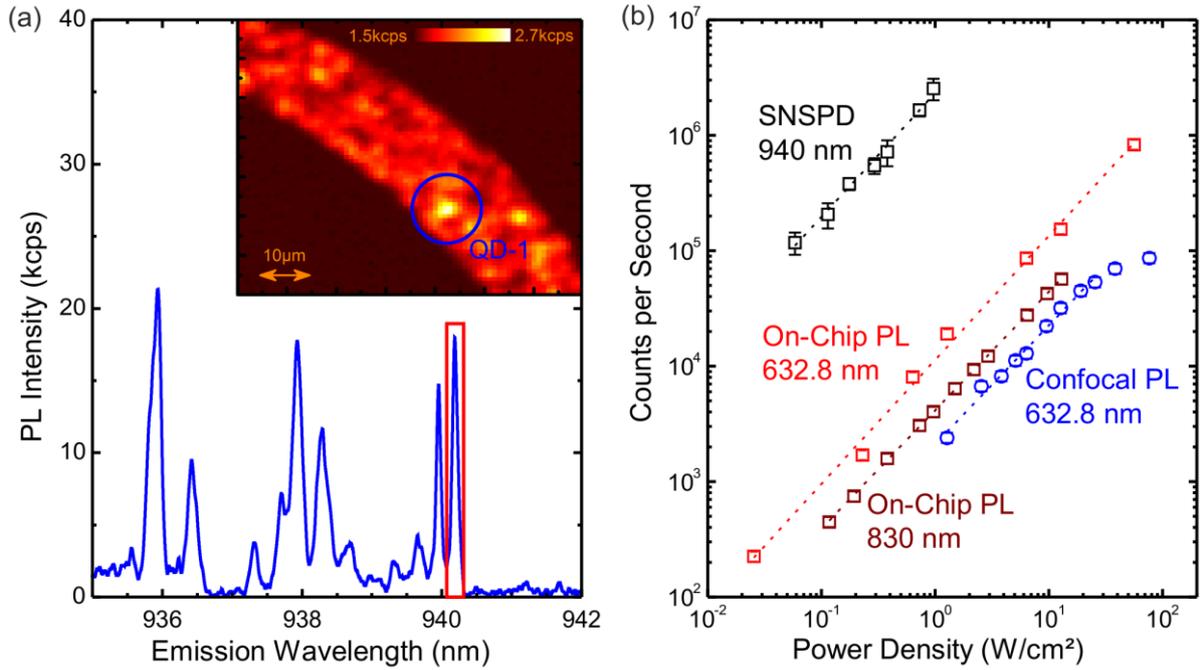

**Figure 4.** (a) Typical quantum dot luminescence spectrum for the position marked QD-1 in the inset. Inset: Spatially resolved quantum dot luminescence detected in confocal geometry for λ = 940.2 ± 0.1 nm (highlighted in red in the spectrum). (b) Detection events as a function of power density for direct SNSPD illumination, as shown in black, and for quantum dot excitation, marked as "On-Chip PL". For comparison the confocal PL of the exciton highlighted in fig 2a is shown in blue.